\documentclass[american,english]{msml2020_arxiv}
\usepackage[T1]{fontenc}
\usepackage[latin9]{inputenc}
\usepackage{amsmath}
\usepackage{amssymb}
\usepackage{graphicx}

\makeatletter
\usepackage{hyperref}

\msmlauthor{
Andreas Mardt\footnotemark[1] \and
Luca Pasquali\footnotemark[1] \\
\addr Freie Universität Berlin, Department of Mathematics and Computer Science, Arnimallee 6, 14195 Berlin, Germany \AND
Frank No\'{e}\Email{frank.noe@fu-berlin.de} \\
\addr Freie Universität Berlin, Department of Mathematics and Computer Science, Arnimallee 6, 14195 Berlin, Germany \\
\addr Freie Universität Berlin, Department of Physics, Arnimallee 6, 14195 Berlin, Germany \\
\addr Rice University, Department of Chemistry, Houston TX, 77005, United States \AND 
Hao Wu \Email{hwu@tongji.edu.cn} \\
\addr Tongji University, School of Mathematical Sciences, Shanghai, 200092, P.R. China
}

\makeatother

\usepackage{babel}
\begin{document}
\title{Deep learning Markov and Koopman models with physical constraints}
\maketitle
\begin{abstract}
The long-timescale behavior of complex dynamical systems can be described
by linear Markov or Koopman models in a suitable latent space. Recent
variational approaches allow the latent space representation and the
linear dynamical model to be optimized via unsupervised machine learning
methods. Incorporation of physical constraints such as time-reversibility
or stochasticity into the dynamical model has been established for
a linear, but not for arbitrarily nonlinear (deep learning) representations
of the latent space. Here we develop theory and methods for deep learning
Markov and Koopman models that can bear such physical constraints.
We prove that the model is an universal approximator for reversible
Markov processes and that it can be optimized with either maximum
likelihood or the variational approach of Markov processes (VAMP).
We demonstrate that the model performs equally well for equilibrium
and systematically better for biased data compared to existing approaches,
thus providing a tool to study the long-timescale processes of dynamical
systems.\foreignlanguage{american}{\begin{keywords}
Markov state model, Koopman model, phsyical constraints, deep learning, dynamical systems
\end{keywords}}
\end{abstract}

\section{Introduction}

\selectlanguage{american}%
\renewcommand*{\footnoteseptext}{ }\renewcommand{\thefootnote}{\fnsymbol{footnote}}\footnotetext[1]{equal contribution}\foreignlanguage{english}{Markovian,
or linear dynamical models are very successful in describing the
effective or long-term dynamics of complex dynamical systems, such
as molecular dynamics (MD) \cite{SchuetteFischerHuisingaDeuflhard_JCompPhys151_146,NoeHorenkeSchutteSmith_JCP07_Metastability,SwopePiteraSuits_JPCB108_6571,ChoderaEtAl_JCP07,PrinzEtAl_JCP10_MSM1,ChoderaNoe_COSB14_MSMs},
wireless communications \cite{konrad2001markov,ma2001composite},
and fluid dynamics \cite{Schmid_JFM10_DMD,mezic2013analysis,TuEtAl_JCD14_ExactDMD,froyland2016optimal}.
The cornerstone of modeling complex nonlinear dynamics with a linear
and often low-dimensional model, such as a Markovian transition matrix,
is that the dynamics can be linearized in the space of eigenfunctions
or singular functions of the corresponding full-space dynamical operator
\cite{SchuetteFischerHuisingaDeuflhard_JCompPhys151_146,Mezic_NonlinDyn05_Koopman,Wu2019}.
In stochastic systems, such a linear model is often called Markov
model, as it is convenient to describe the long-time dynamics as a
Markov chain or Markov jump process between discrete states, whereas
it is often called Koopman model in complex dynamical systems analysis
or fluid dynamics. Markov models and Koopman models greatly simplify
analysis of the dynamical systems when compared to models with explicit
memory terms.}

\selectlanguage{english}%
A wide variety of such linear dynamical models has been developed
across different fields, including Markov State Models (MSMs) \cite{SchuetteFischerHuisingaDeuflhard_JCompPhys151_146,PrinzEtAl_JCP10_MSM1,BowmanPandeNoe_MSMBook},
Markov transition models \cite{WuNoe_JCP15_GMTM}, Ulam's Galerkin
method \cite{dellnitz2001algorithms,bollt2013applied,froyland2014computational},
blind-source separation \cite{Molgedey_94,ZieheMueller_ICANN98_TDSEP},
the variational approach for conformational dynamics (VAC) \cite{NoeNueske_MMS13_VariationalApproach,NueskeEtAl_JCTC14_Variational},
time-lagged independent component analysis (TICA) \cite{PerezEtAl_JCP13_TICA,SchwantesPande_JCTC13_TICA},
dynamic mode decomposition (DMD) \cite{RowleyEtAl_JFM09_DMDSpectral,Schmid_JFM10_DMD,TuEtAl_JCD14_ExactDMD},
extended dynamic mode decomposition (EDMD) \cite{WilliamsKevrekidisRowley_JNS15_EDMD},
variational Koopman models \cite{WuEtAl_JCP17_VariationalKoopman},
variational diffusion maps \cite{BoninsegnaEtAl_JCTC15_VariationalDM},
kinetic maps \cite{NoeClementi_JCTC15_KineticMap,NoeClementi_JCTC16_KineticMap2},
the variational approach of Markov processes (VAMP) \cite{WuEtAl_JCP17_VariationalKoopman},
sparse identification of nonlinear dynamics \cite{brunton2016discovering}
and corresponding kernel embeddings \cite{HarmelingEtAl_NeurComput03_KernelTDSEP,SchwantesPande_JCTC15_kTICA,song2013kernel}
and tensor formulations \cite{NueskeEtAl_JCP15_Tensor,KlusSchuette_Arxiv15_Tensor}.
All these models approximate the Markov dynamics through a linear
model:
\begin{equation}
\mathbb{E}[(\mathbf{g}(\mathbf{x}_{t+\tau})]=\mathbf{K}^{T}\mathbb{E}[\mathbf{f}(\mathbf{x}_{t})]\label{eq:linear_model}
\end{equation}
where \textbf{$\mathbf{f},\mathbf{g}$ }transform the configuration
$\mathbf{x}$ into a latent space representation in which the dynamics
are linear, usually the space of slow transitions or rare events \cite{NoeClementi_COSB17_SlowCVs}.
\textbf{$\mathbf{K}$} is an MSM transition matrix, or a Koopman model,
and can be interpreted as a finite-rank approximation of the full-dimensional
dynamical Markov operator \cite{SchuetteFischerHuisingaDeuflhard_JCompPhys151_146,Koopman_PNAS31_Koopman,Mezic_NonlinDyn05_Koopman,Wu2019}.

MSMs have been particularly successful as methods to extract the long-timescale
kinetics from high-throughput MD simulation data, e.g. of protein
dynamics \cite{Bowman_JCP09_Villin,PrinzEtAl_JCP10_MSM1,BuchFabritiis_PNAS11_Binding,ShuklaPande_NatCommun14_SrcKinase,SilvaEtAl_PNAS14_RNAPolymeraseII,ReuboldEtAl_Nature15_DynaminTetramer,PlattnerEtAl_NatChem17_BarBar}.
In recent years, the MD community has seen a rapid increase in the
available amount of simulated data of complex molecular systems due
to advances in both computing power and simulation techniques \cite{LindorffLarsenEtAl_Science11_AntonFolding,PlattnerEtAl_NatChem17_BarBar,KohlhoffEtAl_NatChem14_GPCR-MSM,DoerrEtAl_JCTC16_HTMD}.
Unlike experiments, MD simulations can resolve structure and dynamics
simultaneously. The extraction of kinetic, i.e. long-timescale information
from simulation data, however, is not trivial, since kinetic information
cannot be inferred from structural similarity \cite{Keller_JCP2010_Clustering,KrivovKarplus_PNAS101_14766,NueskeEtAl_JCTC14_Variational},
as similar structures may be separated by high energy barriers.

Whereas MSM construction has previously been a relatively complex
pipeline of feature selection, dimension reduction, clustering, estimating
the transition matrix $\mathbf{K}$, etc, these choices have recently
been guided by variational approaches \cite{NoeNueske_MMS13_VariationalApproach,NueskeEtAl_JCTC14_Variational,SchwantesPande_JCTC13_TICA,PerezEtAl_JCP13_TICA,Wu2019,MardtEtAl_VAMPnets,Chen_2019}.
These variational methods aim to approximate the leading eigenfunctions
or singular functions of the Markov process, which parametrize the
long-time kinetics, and in which approximately linear Markovian models
(\ref{eq:linear_model}) can be obtained. \cite{MardtEtAl_VAMPnets}
recently proposed VAMPnet that simultaneously learns neural networks
for the latent space representations $\mathbf{f}$ and $\mathbf{g}$
as well as the transition matrix $\mathbf{K}$ in a single end-to-end
learning framework. VAMPnet uses the variational approach of Markov
processes (VAMP) \cite{Wu2019} in order to optimally find $\mathbf{f}$
and $\mathbf{g}$. The VAMPnet framework has been further developed,
for example for learning directly the eigenfunctions of the Markov
operator rather than the MSM transition matrix \cite{Chen_2019},
and for transferring parameters across chemical space \cite{xie2019graph}.

By means of their ability to represent very nonlinear latent space
representations, VAMPnets have been demonstrated to learn high-quality
MSMs with little input from human experts \cite{MardtEtAl_VAMPnets}.
However, there is an important aspect that is well established with
``shallow'' manually constructed MSMs and is yet unsolved with deep
learning methods for MSMs such as VAMPnets: the incorporation of physical
constraints into the transition matrix $\mathbf{K}$, especially reversibility
(detailed balance) and stochasticity.

A dynamical system is statistically reversible when the absolute (unconditional)
probability of finding a transition from point $\mathbf{x}$ at time
$t$ to point $\mathbf{y}$ at time $t+\tau$ is equal to the reverse.
Physically, this occurs when the system, e.g. molecule is simulated
in equilibrium, i.e. without applying external forces. Consequences
of statistical reversibility are that (i) there is no probability
flux in cycles, consistent with the second law of thermodynamics stating
that no work can be extracted from a system purely driven by thermal
energy, (ii) $\mathbf{K}$ can be symmetrized with an equilibrium
vector and thus (iii) the transition matrix $\mathbf{K}$ has real
eigenvalues. If $\mathbf{K}$ is a stochastic transition matrix, this
means that it fulfills detailed balance with respect to its equilibrium
vector. Even if the underlying MD system has been simulated in equilibrium,
estimating a VAMPnet from finite data will not guarantee a reversible
model $\mathbf{K}$. For standard MSMs, maximum likelihood estimators
have been developed to enforce reversibility \cite{Noe_JCP08_TSampling,Bowman_JCP09_Villin,TrendelkampNoe_JCP13_EfficientSampler,TrendelkampSchroerEtAl_InPrep_revMSM}.
\cite{Chen_2019} approached the problem by symmetrizing the covariance
matrices involved in the estimation of $\mathbf{K}$, which works
well for long simulation trajectories, but introduces significant
bias for many short trajectories emerging from a non-equilibrium distribution.
As indicated in \cite{MardtEtAl_VAMPnets}, reversible VAMPnets could
be developed by using the Koopman reweighting method proposed in \cite{WuEtAl_JCP17_VariationalKoopman},
however that approach reduces the bias at a cost of a quite large
estimator variance. Here we develop a new deep learning framework
for Markov processes that can learn reversible transition models from
non-equilibrium data in a robust and end-to-end manner.

Furthermore, a transition matrix $\mathbf{K}$ estimated via a VAMPnet
is not automatically a stochastic matrix. Even when the VAMPnet encoders
map to a state assignment, e.g. by using a SoftMax output \cite{MardtEtAl_VAMPnets},
the transition matrix will have probability mass conservation, i.e.
row sums equal 1, but may still have negative elements. With such
a Markov model we can compute valid observables, such as propagated
probability vectors, a valid equilibrium distribution and meaningful
correlation functions. But the individual matrix elements of $\mathbf{K}$
can no longer be interpreted as transition probabilities or rates
and therefore some analyses, such as transition path theory \cite{EVandenEijnden_TPT_JStatPhys06,MetznerSchuetteVandenEijnden_TPT,NoeSchuetteReichWeikl_PNAS09_TPT},
are no longer applicable. The deep learning framework for Markov processes
introduced here can optionally enforce the $\mathbf{K}$ matrix to
be a stochastic matrix.

In summary, our \textbf{contributions} are as follows:
\begin{enumerate}
\item We develop a flexible deep learning framework for Markov processes,
in which deep networks can be used to learn the latent space representation
of the system and a linear model $\mathbf{K}$ is learned to describe
the time propagation in latent space.
\item We develop a way to enforce $\mathbf{K}$ to be reversible or non-reversible,
optionally. We proove that when enforcing reversibility we obtain
a universal approximator for reversible Markov processes.
\item We develop a way to optionally enforce $\mathbf{K}$ to be stochastic,
i.e. have nonnegative elements. We can combine both physical constraints
in either way, obtaining four choosable combinations of reversible/non-reversible
and stochastic/non-stochastic VAMPnets or deep MSMs.
\item We provide two optimization targets for our method, using maximum
likelihood or VAMP as a loss function, and demonstrate that they provide
asymptotically unbiased results.
\item We illustrate our learning method on benchmark data and demonstrate
that it yields accurate estimates in both the limit of single long
simulation trajectories and many short trajectories, whereas other
methods fail in the latter case.
\end{enumerate}

\section{Theory}

\subsection{Markov processes, spectral decomposition and Koopman theory}

The dynamics of a Markovian dynamical system can be modeled by the
transition density, i.e. the probability density to transition to
a state space point $\mathbf{y}$ at time $t+\tau$, given that the
system was at state $\mathbf{x}$ at time $t$:
\[
p_{\tau}(\mathbf{x},\mathbf{y})=\mathbb{P}(\mathbf{x}_{t+\tau}=\mathbf{y}\mid\mathbf{x}_{t}=\mathbf{x}).
\]
Based on the transition density, we can characterize the time evolution
of the ensemble of system states as
\[
p_{t+\tau}(\mathbf{y})=\left(\mathcal{P}_{\tau}p_{t}\right)(\mathbf{y})\triangleq\int p_{\tau}(\mathbf{x},\mathbf{y})\:p(\mathbf{x})\:\mathrm{d}\mathbf{x}
\]
where $p_{t}$ is the probability density of the system being in any
state at time $t$ and the lag time $\tau$ is the time resolution
of the model. The propagation of general observable functions $f$
can be modeled as
\[
\mathbb{E}[f(\mathbf{x}_{t+\tau})|\mathbf{x}_{t}=\mathbf{x}]=\left(\mathcal{K}_{\tau}f\right)(\mathbf{x})\triangleq\int p_{\tau}(\mathbf{x},\mathbf{y})\:f(\mathbf{y})\:\mathrm{d}\mathbf{y}.
\]
The integral operators $\mathcal{P_{\tau}}$ and $\mathcal{K}_{\tau}$
are called \emph{propagator} and \emph{Koopman operator} respectively,
and are both able to fully describe the Markovian dynamics. From hereon,
we only consider the Koopman operator based modeling formalism, which
is commonly used in the field of dynamical systems (see e.g., \cite{Mezic_NonlinDyn05_Koopman}),
but all conclusions in this paper can be equivalently established
by using the propagator description.

The Koopman operator is linear but infinite-dimensional, and we can
generally approximate the essential part of the dynamics at long timescales
by a finite-dimensional linear model in the form of Eq. (\ref{eq:linear_model})
with $\mathbf{f}$ and $\mathbf{g}$ being two sets of latent variables.
Denoting by $\rho_{0}$ the empirical distribution of $\mathbf{x}_{t}$
and $\rho_{1}$ the empirical distribution of $\mathbf{x}_{t+\tau}$
in all transition pairs $(\mathbf{x}_{t},\mathbf{x}_{t+\tau})$, it
can be proven that an optimal finite-rank approximation of the transition
density can be written in the form
\begin{equation}
\hat{p}_{\tau}(\mathbf{x},\mathbf{y})=\mathbf{f}(\mathbf{x})^{\top}\mathbf{S}\mathbf{g}(\mathbf{y})\rho_{1}(\mathbf{y}),\label{eq:low_rank_general}
\end{equation}
where $\mathbf{S}=\mathbf{K}\left(\mathbb{E}[\mathbf{g}(x_{t+\tau})\mathbf{g}(x_{t+\tau})^{T}]\right)^{-1}$
\cite{Wu2019}. Based on conformation dynamics theory \cite{SchuetteFischerHuisingaDeuflhard_JCompPhys151_146},
Koopman theory \cite{Mezic_NonlinDyn05_Koopman,Koopman_PNAS31_Koopman},
and the variational approach of Markov processes (VAMP) \cite{Wu2019},
the finite-dimensional model can accurately capture the essential
or long-time part of the dynamics by selecting $\mathbf{f},\mathbf{g}$
to be the dominant singular functions or eigenfunctions of the Koopman
operator. Specifically, if we consider the modeling error of the Koopman
operator in the sense of Hilbert-Schmidt norm, the optimal model can
be given by the truncated singular value decomposition of the transition
density 
\begin{equation}
p_{\tau}(\mathbf{x},\mathbf{y})\approx\sum_{i=1}^{k}\sigma_{i}\psi_{i}(\mathbf{x})\phi_{i}(\mathbf{y})\rho_{1}(\mathbf{y}),\label{eq:finite-dimensional-transition-density}
\end{equation}
where $(\sigma_{1},\ldots,\sigma_{k})$ are the largest singular values
of the Koopman operator $\mathcal{K}_{\tau}$, $(\psi_{1},\ldots,\psi_{k})$
and $(\phi_{1},\ldots,\phi_{k})$ are the corresponding dominant left
and right singular functions, and the equality holds exactly if $k\to\infty$.

If we further assume that the dynamics are statistically reversible,
which implies that the system does not contain net cycles and there
is no work produced in equilibrium, then the detailed balance condition
\begin{equation}
\mu(\mathbf{x})p_{\tau}(\mathbf{x},\mathbf{y})=\mu(\mathbf{y})p_{\tau}(\mathbf{y},\mathbf{x}),\label{eq:detailed_balance}
\end{equation}
is satisfied, where $\mu(\mathbf{x})$ is the equilibrium distribution
of system states. This means that the unconditional probability to
observe the transition $\mathbf{x}\rightarrow\mathbf{y}$ is equal
to that of transition $\mathbf{y}\rightarrow\mathbf{x}$. In this
case, the Koopman operator is a self-adjoint operator, and the truncated
singular value decomposition of the dynamics is equivalent to the
truncated eigendecomposition
\[
p_{\tau}(\mathbf{x},\mathbf{y})\approx\sum_{i=1}^{k}\lambda_{i}\varphi_{i}(\mathbf{x})\varphi_{i}(\mathbf{y})\mu(\mathbf{y})
\]
with eigenfunctions $\varphi_{i}=\psi_{i}=\phi_{i}$ and eigenvalues
$\lambda_{i}=\sigma_{i}$ when the equilibrium is achieved by data.

The eigenvalues and eigenfunctions can be systematically approximated
from data by invoking the variational approach of conformation dynamics
(VAC) \cite{NueskeEtAl_JCTC14_Variational,MardtEtAl_VAMPnets,Wu2019},
which provides a loss function with which hyper-parameter selection
can be made in the traditional MSM construction pipeline \cite{McGibbonPande_JCP15_CrossValidation},
and which can be used in order to train a deep neural network to represent
the eigenfunctions \cite{Chen_2019}. Likewise, the singular values
and singular functions of non-reversible or even non-stationary dynamical
models can be variationally approximated with VAMP \cite{Wu2019},
which in turn can be used for hyper-parameter selection in MSM pipelines
\cite{scherer_husic_variational_selection} and deep learning molecular
kinetics with VAMPnets \cite{MardtEtAl_VAMPnets}.

In this paper, we will also use neural networks as universal function
approximators to find latent variables with a suitable optimization
principle. The main novelty is that our framework allows to obtain
a finite-dimensional model that incorporates desired physical constraints
such as stochasticity and reversibility in an asymptotically unbiased
way.

\section{Deep Markov Model with physical constraints}

\subsection{Transition Model}

\begin{figure}
\centering\includegraphics[width=0.6\textwidth]{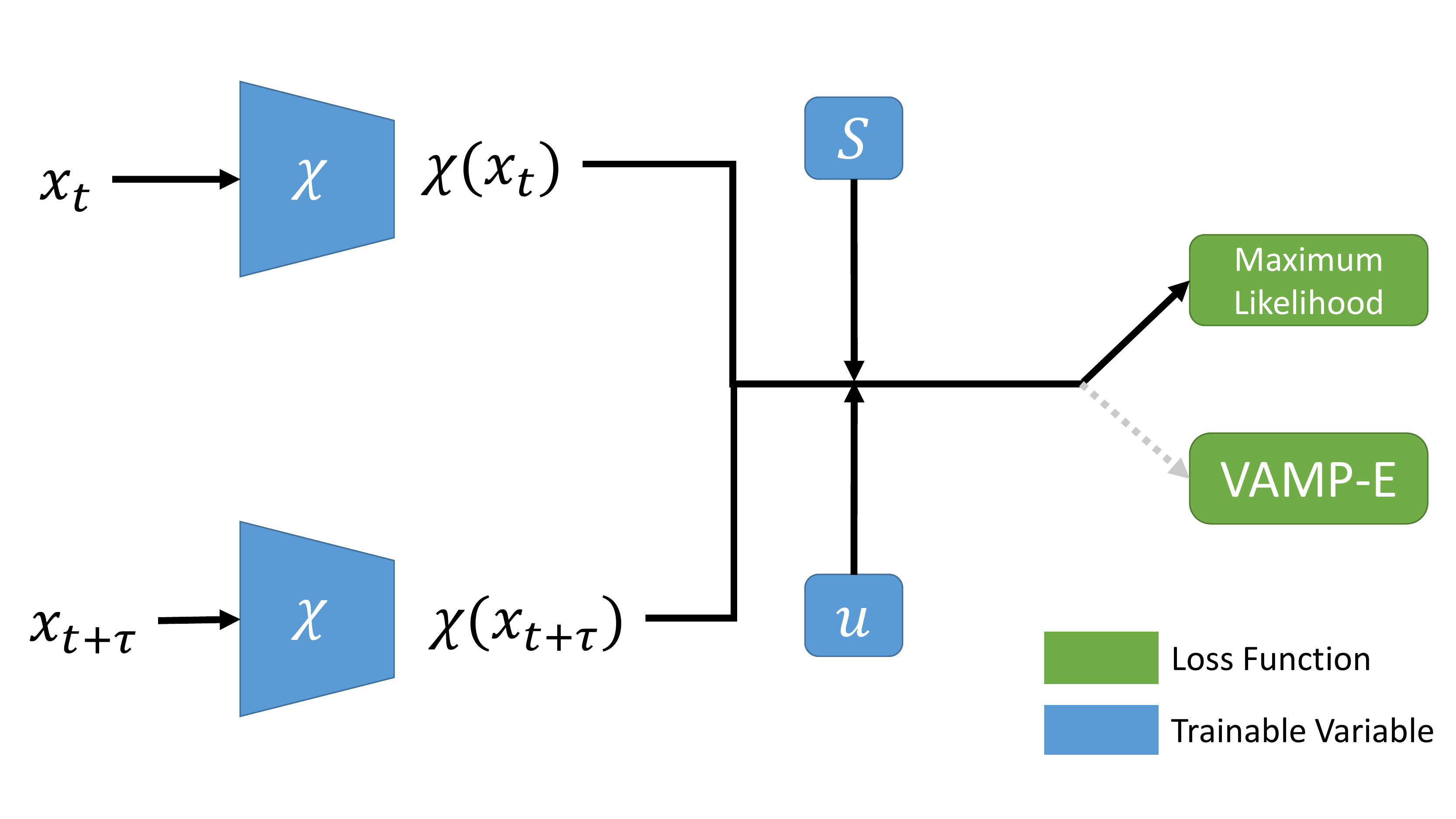}

\caption{\textbf{Schematic of the proposed deep learning architecture for Markov
processes with physical constraints}. For each time step $t$ of the
training data, the coordinates $\mathbf{x}_{t}$ and $\mathbf{x}_{t+\tau}$
are passed to two instances of the same network $\boldsymbol{\chi}$.
The transformed outputs are then concatenated, and used together with
the trainable variables $\mathbf{S},\mathbf{u}$ to obtain an estimate
of the transition density $p(\mathbf{y}=\mathbf{x}_{t+\tau}\mid\mathbf{x}=\mathbf{x}_{t}).$
This probability estimate allows us to train $\boldsymbol{\chi},\mathbf{S},\mathbf{u}$
using a Maximum Likelihood loss. Alternatively, the whole architecture
can be trained using as an optimization target the VAMP-E score. The
scheme is similar to a classical VAMPnet \cite{MardtEtAl_VAMPnets},
but has the ability to build in physical constraints such as reversibility
and stochasticity into the parameters $\mathbf{S},\mathbf{u}$. \label{fig:figure_network_structure}}
\end{figure}

The aim of this section is to construct a general model which approximates
the Markovian process defined by the transition density; this density
can be learned with neural networks and physical constraints for reversibility
and stochasticity of the transition matrix can be built into it. We
propose the following model of the transition density:
\begin{equation}
p_{\tau}(\mathbf{x},\mathbf{y})=\boldsymbol{\chi}(\mathbf{x})^{T}\mathbf{S}\boldsymbol{\chi}(\mathbf{y})\boldsymbol{\chi}(\mathbf{y})^{T}\mathbf{u}\rho_{1}(\mathbf{y}),\label{eq:Markovdensity}
\end{equation}
which is a special representation of the general finite-dimensional
model (\ref{eq:low_rank_general}-\ref{eq:finite-dimensional-transition-density})
with the choices $\mathbf{f}(\mathbf{x})=\boldsymbol{\chi}(\mathbf{x})$
and $\mathbf{g}(\mathbf{y})=\boldsymbol{\chi}(\mathbf{y})\boldsymbol{\chi}(\mathbf{y})^{T}\mathbf{u}$.
However, we will demonstrate that this choice leads to an universal
approximator for the Markov processes of interest. Here, $\boldsymbol{\chi}$
is a neural network that maps a sample $\mathbf{x}$ from configuration
space onto a fuzzy clustering, which can be thought of indicating
to what degree $\mathbf{x}$ belongs to each of the few metastable
states. The trainable vector $\mathbf{u}$ is necessary to reweight
the empirical distribution towards the equilibrium distribution of
the system. Eq. (\ref{eq:Markovdensity}) can be understood as mapping
the starting sample $\mathbf{x}$ to the state representation $\boldsymbol{\chi}(\mathbf{x})$,
making a time propagation with $\mathbf{S}$, and checking how close
it ends up to the state representation $\boldsymbol{\chi}(\mathbf{y})$,
which is weighted with respect to the system's equilibrium probability
(Fig. \ref{fig:figure_network_structure}).

Here we work with stationary dynamical systems and thus assume that
there exists a unique equilibrium distribution $\mu(\mathbf{x})=\mathbf{u}^{T}\boldsymbol{\chi}(\mathbf{x})\rho(\mathbf{x})$.
Hence our model needs to always fulfill the following constraints
to guarantee a normalized equilibrium and transition density (see
Appendix \ref{subsec:Proof-of-normalization} for proof):
\begin{enumerate}
\item \textbf{Normalization of state assignment}: $\boldsymbol{\chi}(\mathbf{x})^{T}\mathbf{1}=1.$
\item \textbf{Normalization of the reweighting vector}: $\bar{\boldsymbol{\chi}}^{T}\mathbf{u}=1$,
where $\bar{\boldsymbol{\chi}}=\mathbb{E}\left[\boldsymbol{\chi}(\mathbf{x}_{t+\tau})\right]$
is the empirical state probability.
\item \textbf{Normalization of transition density}: $\mathbf{S}\mathbf{C}_{\tau\tau}^{\prime}\mathbf{u}=\mathbf{1}$
where $\mathbf{C}_{\tau\tau}^{\prime}=\mathbb{E}\left[\boldsymbol{\chi}(\mathbf{x}_{t+\tau})\boldsymbol{\chi}(\mathbf{x}_{t+\tau})^{\top}\right]$
is the empirical covariance matrix of $\boldsymbol{\chi}(\mathbf{x}_{t+\tau})$.
As a result the Koopman matrix preserves probability mass by means
of $\mathbf{K1}=\mathbf{\mathbf{S}\mathbf{C}_{\tau\tau}^{\prime}\mathbf{u}=\mathbf{1}}$.
\end{enumerate}
By substituting the equilibrium distribution $\mu(\mathbf{x})$ into
(\ref{eq:Markovdensity}), the finite-dimensional model proposed in
this section can be rewritten as
\begin{equation}
p_{\tau}(\mathbf{x},\mathbf{y})=\boldsymbol{\chi}(\mathbf{x})^{T}\mathbf{S}\boldsymbol{\chi}(\mathbf{y})\mu(\mathbf{y}).\label{eq:Markovdensity-1}
\end{equation}
So the equilibrium Koopman matrix $\mathbf{K}$ acts as:
\[
\mathbb{E}[\boldsymbol{\chi}(\mathbf{x}_{t+\tau})]=\mathbf{K}^{\top}\mathbb{E}[\boldsymbol{\chi}(\mathbf{x}_{t})],
\]
and can be estimated from:
\begin{align}
\mathbf{K} & =\mathbf{S\boldsymbol{\Sigma}}\label{eq:Koopman_matrix_calculation}
\end{align}
with
\begin{align}
\boldsymbol{\Sigma} & =\int\boldsymbol{\chi}(\mathbf{y})\mu(\mathbf{y})\boldsymbol{\chi}(\mathbf{y})^{T}\mathrm{d}\mathbf{y}\nonumber \\
 & =\int\boldsymbol{\chi}(\mathbf{y})\rho_{1}(\mathbf{y})\boldsymbol{\chi}(\mathbf{y})^{T}\mathbf{u}\boldsymbol{\chi}(\mathbf{y})^{T}\mathrm{d}\mathbf{y}\label{eq:Sigma}
\end{align}
being the equilibrium covariance matrix of $\boldsymbol{\chi}$. See
Appendix \ref{subsec:Relationships-to-traditional} for an overview
of the relationship with traditional MSMs.

\subsection{Reversible deep Markov Model}

In addition to the necessary constraints introduced above, we have
the choice of enforcing further physical constraints. In order to
enforce detailed balance (\ref{eq:detailed_balance}), or statistical
reversibility, into the model, we enforce:
\begin{equation}
\mathbf{S}=\mathbf{S}^{T}\label{eq:reversiblity}
\end{equation}

\paragraph{Proof of reversibility}

Inserting (\ref{eq:reversiblity}) into (\ref{eq:Markovdensity})
leads to the detailed balance equation \ref{eq:detailed_balance}.
\begin{align*}
\mu(\mathbf{x})p_{\tau}(\mathbf{x},\mathbf{y}) & =\mathbf{u}^{T}\boldsymbol{\chi}(\mathbf{x})\rho(\mathbf{x})\boldsymbol{\chi}(\mathbf{x})^{T}\mathbf{S}\boldsymbol{\chi}(\mathbf{y})\rho(\mathbf{y})\boldsymbol{\chi}(\mathbf{y})^{T}\mathbf{u}\\
 & =\mu(\mathbf{y})p_{\tau}(\mathbf{y},\mathbf{x}).
\end{align*}
Therefore, the Markov process defined by $p_{\tau}(\mathbf{x},\mathbf{y})$
is a reversible Markov process with stationary distribution $\mu(\mathbf{x})$.

\subsection{Reversible deep MSM}

The mandatory constraints above already enforce $\sum_{j}k_{ij}=1$
for all $i$ with $k_{ij}$ denoting the $(i,j)$-th element of $\mathbf{K}$,
and hence probability mass conservation. In order to force the Koopman
matrix $\mathbf{K}$ to be a stochastic Markov chain matrix, we additionally
need to achieve non-negative elements:
\begin{align}
k_{ij} & \ge0\:\:\:\:\forall i,j\label{eq:nonnegativity}
\end{align}
This can be enforced by the additional constraints:
\begin{equation}
\text{All elements of }\mathbf{u},\mathbf{S},\text{ and }\boldsymbol{\chi}\text{ are non-negative.}\label{eq:non-negative}
\end{equation}
The non-negativity (\ref{eq:nonnegativity}) directly follows from
(\ref{eq:Koopman_matrix_calculation}-\ref{eq:Sigma}), since all
factors are positive.

Additionally, constraint (\ref{eq:non-negative}) ensures that the
full transition density is a real probability density, i.e. $p_{\tau}(\mathbf{x},\mathbf{y})\geq0$
$\forall\mathbf{x},\mathbf{y}$. However, this condition may be violated
when using a finite rank approximation of a Koopman model \cite{Wu2019}.

\subsection{Choosing physical constraints}

By toggling the two optional physical constraints, reversibility and
stochasticity, we can change the class of Markov Model we want to
build:
\begin{enumerate}
\item \textbf{Non-reversible VAMPnet} (VAMPnet): This is the most general
case, where both optional constraints are not enforced. This grants
the model the highest flexibility to approximate the Koopman operator.
Hence the model will obtain the best approximation of the eigenfunctions
of the Koopman operator. It can be understood as an alternative approach
to VAMPnets \cite{MardtEtAl_VAMPnets}, where the proposed model learns
a reweighting vector and the Koopman matrix via additional parameters.
\item \textbf{Non-reversible deep MSM} (DMSM): If we obey the non-negativity
constraints (Eq. \ref{eq:non-negative}), we obtain a model with a
stochastic transition matrix $\mathbf{K}$ and a nonnegative transition
density $p_{\tau}(\mathbf{x},\mathbf{y})$. A DMSM can be viewed as
a special case of the general deep MSM \cite{wu2018deep} with $\mathbf{q}(\mathbf{y})=\mathbf{S}\boldsymbol{\chi}(\mathbf{y})\rho(\mathbf{y})\boldsymbol{\chi}(\mathbf{y})^{T}\mathbf{u},$
where $q_{i}(\mathbf{y})$ is the probability of jumping to configuration
$\mathbf{y}$ when starting from state $i$.
\item \textbf{Reversible VAMPnet} (RevVAMPnet): Activating the reversibility
constraint (Eq. \ref{eq:reversiblity}) results in a reversible transition
density with respect to the equilibrium distribution $\mu$.
\item \textbf{Reversible deep MSM} (RevDMSM): The last model combines both
constraints (Eq. \ref{eq:reversiblity} \& \ref{eq:non-negative})
to ensure a reversible model and a stochastic transition matrix. Such
a model is desirable when seeking a reversible dynamical model that
should be analyzed with algorithms operating on individual transition
probabilities, such as transition path theory or committor analyses.
To the best knowledge of the authors no algorithm exists to train
a neural network framework which results in such a model.
\end{enumerate}
Since for the first two models results are already reported, we will
focus in our analysis on the latter two models.

\paragraph{Symmetrized VAMPnet}

(SymVAMPnet)

We compare our reversible models to a previously proposed model \cite{Chen_2019,WuEtAl_JCP17_VariationalKoopman},
which estimates a VAMPnet but additionally enforces reversibility
by symmetrizing the correlation matrices entering the VAMP score as
follows:
\begin{align*}
\mathbf{C}_{00} & =\frac{1}{2}(\mathbb{E}[\boldsymbol{\chi}(\mathbf{x}_{t})\boldsymbol{\chi}(\mathbf{x}_{t})^{T}]+\mathbb{E}[\boldsymbol{\chi}(\mathbf{x}_{t+\tau})\boldsymbol{\chi}(\mathbf{x}_{t+\tau})^{T}]),\\
\mathbf{C}_{01} & =\frac{1}{2}(\mathbb{E}[\boldsymbol{\chi}(\mathbf{x}_{t})\boldsymbol{\chi}(\mathbf{x}_{t+\tau})^{T}]+\mathbb{E}[\boldsymbol{\chi}(\mathbf{x}_{t+\tau})\boldsymbol{\chi}(\mathbf{x}_{t})^{T}]),\\
\mathbf{C}_{11} & =\mathbf{C}_{00}.
\end{align*}
In the limit of a long equilibrium simulation, this model is asymptotically
unbiased, but it can be subject to a strong bias in the case of short
simulations starting from a non-equilibrium distribution, which is
the main application scenario of Markov modeling (see also \cite{WuEtAl_JCP17_VariationalKoopman}).

\section{Machine Learning Architecture and Algorithm}

\subsection{Loss function}

For training the model two learning objectives can be used, the VAMP-E
score \cite{Wu2019} and maximum likelihood (ML), respectively. In
the case of ML we can directly estimate the likelihood to observe
all the data pairs $(\mathbf{x}_{t},\mathbf{x}_{t+\tau})$ in the
trajectory according to the transition density (\ref{eq:Markovdensity}):
\[
\log(L)=\log(\prod_{t=1}^{T-\tau}p_{\tau}(\mathbf{x}_{t},\mathbf{x}_{t+\tau}))=\sum_{t=1}^{T-\tau}\log(p_{\tau}(\mathbf{x}_{t},\mathbf{x}_{t+\tau})).
\]
In order to use ML training, it is necessary that $p_{\tau}(\mathbf{x},\mathbf{y})\geq0$,
which is exclusively the case for the non-reversible and reversible
MSM, i.e. when enforcing constraint (\ref{eq:non-negative}).

If we consider the VAMP-E score, we can rewrite:
\[
p_{\tau}(\mathbf{x},\mathbf{y})=\boldsymbol{\chi}(\mathbf{x})^{T}\mathbf{S}\boldsymbol{\gamma}(\mathbf{y})\rho(\mathbf{y}),
\]
with
\[
\boldsymbol{\gamma}(\mathbf{x})=\boldsymbol{\chi}(\mathbf{x})\boldsymbol{\chi}(\mathbf{x})^{T}\mathbf{u},
\]
which is a weighted state representation compensating for non-equilibrium
data. The corresponding VAMP-E score to be maximized is then
\[
R=\mathrm{tr}[\mathbf{S}^{T}\mathbf{C_{\chi\chi}\mathbf{S}}\mathbf{C}_{\gamma\gamma}-2\mathbf{S}^{T}\mathbf{C}_{\chi\gamma}],
\]
where
\begin{align*}
\mathbf{C}_{\chi\chi} & =\mathbb{E}[\boldsymbol{\chi}(\mathbf{x}_{t})\boldsymbol{\chi}(\mathbf{x}_{t})^{T}],\\
\mathbf{C}_{\chi\gamma} & =\mathbb{E}[\boldsymbol{\chi}(\mathbf{x}_{t})\boldsymbol{\gamma}(\mathbf{x}_{t+\tau})^{T}],\\
\mathbf{C}_{\gamma\gamma} & =\mathbb{E}[\boldsymbol{\gamma}(\mathbf{x}_{t+\tau})\boldsymbol{\gamma}(\mathbf{x}_{t+\tau})^{T}].
\end{align*}
This score can be employed for all four different model classes and
is the only score we consider to train the Koopman models.

\subsection{Enforcing physical constraints}

Using either loss function, we have to fulfill the constraints by
parameterizing $\mathbf{S}$ and $\mathbf{u}$. First we consider
$\mathbf{u}$. If we choose arbitrary weights $\mathbf{w}^{\text{u}}$,
we can enforce the non-negativity by squeezing the weights through
an exponential and enforce $\bar{\boldsymbol{\chi}}^{\top}\mathbf{u}=1$
by proper normalization:
\[
\mathbf{u}=\frac{\exp(\mathbf{w}^{\text{u}})}{\bar{\boldsymbol{\chi}}^{\top}\exp(\mathbf{w}^{\text{u}})}.
\]
For $\mathbf{S}$ we have arbitrary weights $\mathbf{W}^{\text{S}}$.
The symmetry and non-negativity are enforced via:
\begin{align*}
\mathbf{W}_{1} & =\sigma(\mathbf{W}^{\text{S}})+\sigma(\mathbf{W}^{\text{S}^{T}})\\
\sigma(x) & =\begin{cases}
\exp(x) & ,\text{ if }x<0\\
x+1 & ,\text{ otherwise}
\end{cases}.
\end{align*}
In addition, we need to take care of $\mathbf{S}\mathbf{C}_{\tau\tau}^{\prime}\mathbf{u}=\mathbf{Sv}=\mathbf{1}$.
In order to not reverse the symmetry, we can still change the diagonal
elements via $\mathbf{w_{2}}$. 
\begin{align*}
\mathbf{S} & =\mathbf{W}_{1}+\text{diag (\ensuremath{\mathbf{w_{2}}})}\\
(\mathbf{Sv})_{i} & =\sum_{k}W_{ik}v_{k}+w_{i}v_{i}\\
\Rightarrow w_{i} & =\frac{1-\sum_{k}W_{ik}v_{k}}{v_{i}}.
\end{align*}
Since the expression could result into negative elements for $\mathbf{S}$,
we need to normalize $\mathbf{W}_{1}$ beforehand, optimally by $||\mathbf{W}_{1}\mathbf{v}||_{\inf}$.
Any function which returns a value larger than the maximum norm can
be used, although it should be differentiable for gradient based optimization
methods; furthermore, the choice strongly influences the training
properties, as if the value is not a good approximation of the maximum
norm $\mathbf{S}$ will be dominated by the diagonal elements and
thus harder to train. All the p-norms fulfill the requirements, and
the higher the order the closer to the maximum norm they will be;
we considered $p=20$ to be an acceptable value.

\subsection{Training algorithm}
\begin{enumerate}
\item Train a VAMPnet with SoftMax output as the $\boldsymbol{\chi}$ function
with VAMP-2
\item Train $\mathbf{S}$ and $\mathbf{u}$ while keeping $\boldsymbol{\chi}$
fixed with VAMP-E score or ML with the whole training set in one batch
\item Train everything with VAMP-E or ML
\item For the implied timescales keep $\boldsymbol{\chi}$ fixed and train
only for $\mathbf{S}$ and $\mathbf{u}$ with the whole training set
in one batch.
\end{enumerate}
The first two steps are included to stabilize the training, but are
in principle not necessary. The training might be further stabilized
by setting $\mathbf{u}$ to its optimal value for a given $\boldsymbol{\chi}$
calculated via a non-reversible Koopman $\mathbf{K_{\mathrm{non}}}$
model and the stationary distribution $\boldsymbol{\pi}=\boldsymbol{\pi}\mathbf{K_{\mathrm{non}}}$
as
\[
\mathbf{u=\mathbf{C_{\chi\chi}^{-1}}}\boldsymbol{\pi},
\]
which can be repeated during the training process, where $\mathbf{u}$
must still fulfill the non-negative constraint.

\section{Deep reversible Markov Models are universal approximators for reversible
Markov processes}

In order to show that our model is flexible enough to approximate
any reversible Markov process, we will proove that the proposed model
is a universal approximator for reversible Markov processes. The non-reversible
case was treated earlier by \cite{MardtEtAl_VAMPnets}, and we therefore
do not focus on it here. (see Appendices \ref{subsec:Proof-of-proposition1}
and \ref{subsec:Proof-of-proposition2} for proofs)

\textbf{Proposition 1. }\textit{For a reversible Markov process $\{\mathbf{x}_{t}\}$
with Koopman operator $\mathcal{K_{\tau}}$, if there are constants
$C_{0},C_{1}$, so that
\begin{align*}
\rho_{0}(\mathbf{x})\mu(\mathbf{x})^{-1} & \le C_{0},\\
\rho_{1}(\mathbf{x})^{-1}\mu(\mathbf{x}) & \le C_{1}
\end{align*}
for any $\mathbf{x}$, the Koopman operator $\mathcal{\hat{K_{\tau}^{*}}}$
of the optimal (d+1)-dimensional model in the form of (\ref{eq:Markovdensity})
with the largest VAMP-E score satisfies
\[
||\mathcal{\hat{K_{\tau}^{*}}-\mathcal{K_{\tau}}||{}_{\mathrm{HS}}^{\mathrm{2}}}\le C_{0}C_{1}\sum_{i=d+1}^{\infty}\lambda_{i}^{2},
\]
where $\mu$ is the stationary density, and $\lambda_{i}$ denotes
the i-th largest eigenvalue of $\mathcal{K_{\tau}}$}.

The following proposition shows that in equilibrium we can identify
the dominant eigencomponents with our model.

\textbf{Proposition 2. }\textit{If $\{\mathbf{x}_{t}\}$ is a reversible
and stationary Markov process with Koopman operator $\mathcal{K_{\tau}}$,
the VAMP-E score $\mathcal{R}_{E}$ of the proposed model with $\mathrm{dim}(\boldsymbol{\chi})=d$
satisfies
\begin{equation}
\mathcal{R}_{E}\le\sum_{i=1}^{d}\lambda_{i}^{2},\label{eq:vampe}
\end{equation}
and the equality can hold in the case of $\mathrm{span}(\chi_{1},...,\chi_{d})=\mathrm{span}(\varphi_{1},...,\varphi_{d})$,
where $\lambda_{i}$ denotes the i-th largest eigenvalue of $\mathcal{K_{\tau}}$
with the corresponding eigenfunction $\varphi_{i}$.}

\section{Results}

\paragraph{Overview}

\begin{figure}
\includegraphics[width=0.9\textwidth]{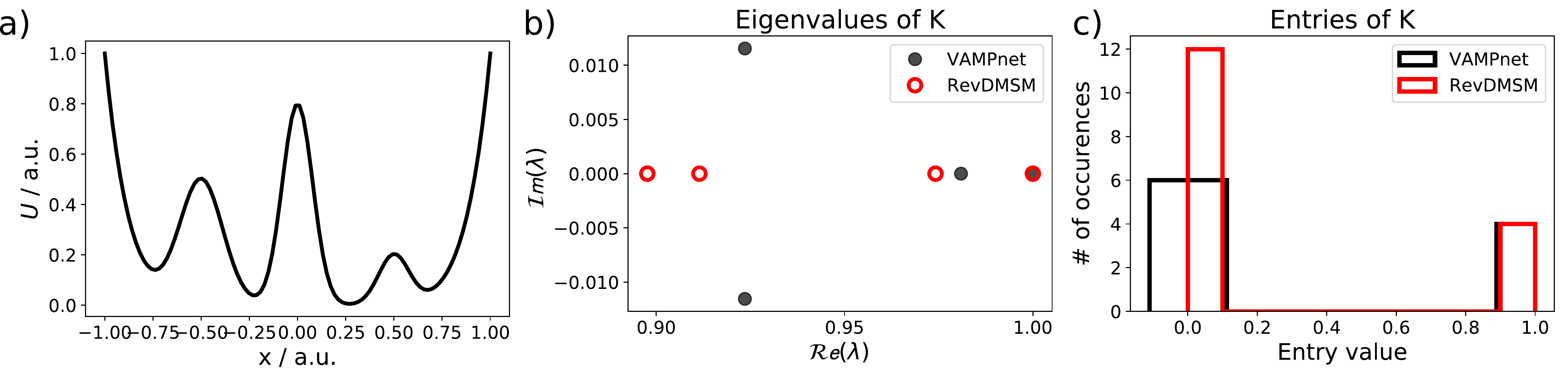}

\caption{\textbf{Demonstration of incorporation of physical constraints: reversibility
and non-negativity}. The eigenvalues and the distribution of elements
of the transition matrix K are shown for an unconstrained VAMPnet
and a RevDMSM trained on poorly sampled training data. a) Potential
energy profile of the Prinz model b) Imaginary and real part of the
eigenvalues of a model estimated with non-reversible VAMPnets and
a RevDMSM. c) Entries of the matrix K of these two models.\label{fig:Figure1}}
\end{figure}

Below we demonstrate our model by applying it to a time-discretized
one-dimensional diffusion process $x_{t+\Delta t}=-\Delta t\nabla V(x_{t})+\sqrt{2\Delta t}\eta_{t}$
in the Prinz potential $V(x)$ \cite{PrinzEtAl_JCP10_MSM1} (Fig.
\ref{fig:Figure1}a) with time step $\Delta t=0.001$ and $\eta_{t}$
being standard normal random variables. When generating training data
we save the state $x$ every five timesteps. The neural network $\boldsymbol{\chi}$
has one input node, receiving the current value of the $x$ coordinate.
We validate that when enforcing the nonnegativity and reversibility
constraints, our models will result in a valid transition matrix and
real eigenvalues respectively, even in the case of poorly sampled
data. Furthermore, we show that our reversible models give unbiased
results for implied timescales and equilibrium probabilities even
when using non-equilibrium data for training, while a simple symmetrization
of the correlation matrices (SymVAMPnet) does not. Finally, we study
the ability of the proposed methods to approximate the exact eigenfunctions
of the test system.

\paragraph{Implementation and training data}

The methods were implemented using \textit{Keras} \cite{chollet2015keras}
with \textit{tensorflow} \cite{tensorflow2015-whitepaper_2} as a
backend.  For the full code and details about the neural network
architecture, hyper-parameters and training routine, please refer
to \href{https://github.com/markovmodel/deep_rev_msm}{https://github.com/markovmodel/deep\_rev\_msm}.

Unless otherwise noted, we used the adam optimizer \cite{adam}, a
batch-size of 5000, and a six-layer-deep neural network with a constant
width of 100 nodes for $\boldsymbol{\chi}$.

As training data we use either a single simulation trajectory of variable
length, or a varying number of short trajectories with fixed length
(see below). Non-equilibrium data are sampled from a starting distribution
with probabilities $[15\%,70\%,9\%,6\%]$ to start at the points $[-0.75+x_{1},-.25+x_{2},.25+x_{3},.75+x_{4}]$
where $x_{i}$ are independent random variables sampled from a Gaussian
distribution with zero mean and standard deviation 0.15.

\paragraph{Reversible VAMPnets and reversible deep MSMs obtain transition matrices
with real eigenvalues and nonnegative entries}

To simulate an insufficiently sampled example, we created 1000 trajectories
of 1 time step with the starting distribution as stated above for
training, validation, and test set respectively. We train a regular
VAMPnet and a reversible Deep Markov State Model (RevDMSM) on the
training data with an early stopping given by the performance on the
validation set and estimate the resulting Koopman matrix on the test
set with a fixed number of output nodes $d=4$. Fig. \ref{fig:Figure1}b)
shows the resulting eigenvalues of these matrices and Fig. \ref{fig:Figure1}c)
the distribution of the entries. Using the non-reversible VAMPnet,
the poor sampling leads to complex eigenvalues and negative entries
for the Koopman matrix. Thus, we not only obtain a non-reversible
model, but the Koopman matrix also does not correspond to a valid
transition matrix. The RevDMSM model does not suffer from these shortcomings,
nevertheless the constraints imposed on the model result in slightly
lower eigenvalues, which can be expected since the constraints hinder
the ability to approximate the eigenfunctions of the Koopman operator.

\paragraph{Reversible VAMPnets converge to unbiased timescales and state probabilities
for biased training data}

\begin{figure}
\includegraphics[width=0.9\textwidth]{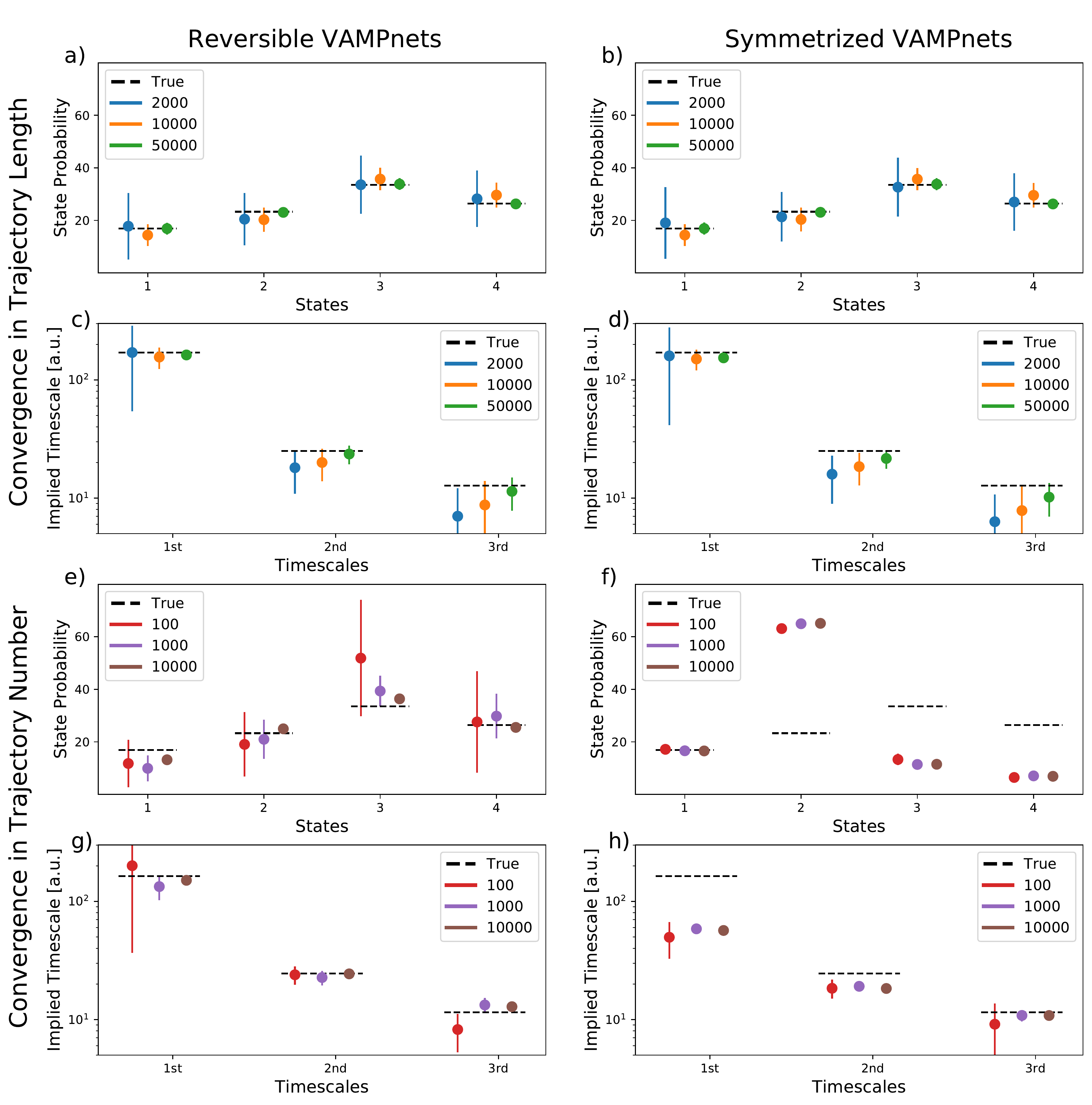}

\caption{\textbf{Reversible VAMPnets converge to unbiased equilibrium probabilities
from biased data}. Comparison of building a Koopman model with a RevVAMPnet
(left column) and SymVAMPnets (right column) on the Prinz potential
dataset with a varying number of trajectories of $11$ frames each
starting from an off-equilibrium distribution (a-d), and varying length
of a single trajectory (e-h). Depicted is the state probability to
be in the four intervals ($[-1,-.5],[-.5,0],[0,.5],[.5,1]$) and the
three slowest timescales as the mean over the lag times $[6,8,10]$,
where the horizontal black line marks the true value (bottom). Errors
are estimated over 5 runs as two sigma intervals.\label{fig:Figure2}}
\end{figure}

In Fig. \ref{fig:Figure2} we compare the performance of RevVAMPnets
and SymVAMPnets when estimating the properties of an poorly sampled
dynamical system, and specifically how well the two methods approximate
the stationary probability of the four main states ($[(-\infty,-.5]$,
$[-0.5,0.]$, $[0.,.5]$, $[.5,\infty)]$) and the estimated values
for the timescales of the dynamical system. We use as benchmark the
Prinz potential, using as training data a varying number of trajectories
with fixed length of $11$ frames, and a single trajectory with a
varying number of frames; we chose a fixed length of $11$ frames
in order to estimate the timescales at $\tau=10$. We test the convergence
over an increasing number of trajectories of the two models using
$10^{2},10^{3},10^{4}$ trajectories, respectively. We also vary the
trajectory length between $2\cdot10^{3},10^{4},5\cdot10^{4}$ frames,
respectively. The true values of the timescales are numerical approximations
by a transition matrix computed for a direct uniform 1000-state discretization
of the $x$-axis for $2\cdot10^{7}$ frames \cite{PrinzEtAl_JCP10_MSM1},
while the true state probabilities were calculated directly from the
analytical expression of the potential.

The test of convergence in trajectory length shows how both methods
converge to the true values of the system's timescales and state probabilities,
as it is expected when the training data distribution converges to
the stationary distribution (Fig. \ref{fig:Figure2} a-d). The test
of convergence in trajectory numbers shows how the RevVAMPnets method
is able to approximate the real state probabilities and timescale
values already with a small number of short trajectories within statistical
uncertainty, and converges to a value consistently close to the real
one when the number of trajectories used as training data increases;
we did not observe this behavior for the SymVAMPnets, as this method
is unable to recover the true dynamics and equilibrium distribution
of the system when working with a heavily biased sampling (Fig. \ref{fig:Figure2}
e-h), which results in a first timescale nearly a factor 3 too low.

\paragraph{Approximation of the true eigenfunctions}

\begin{figure}
\includegraphics[width=0.85\textwidth]{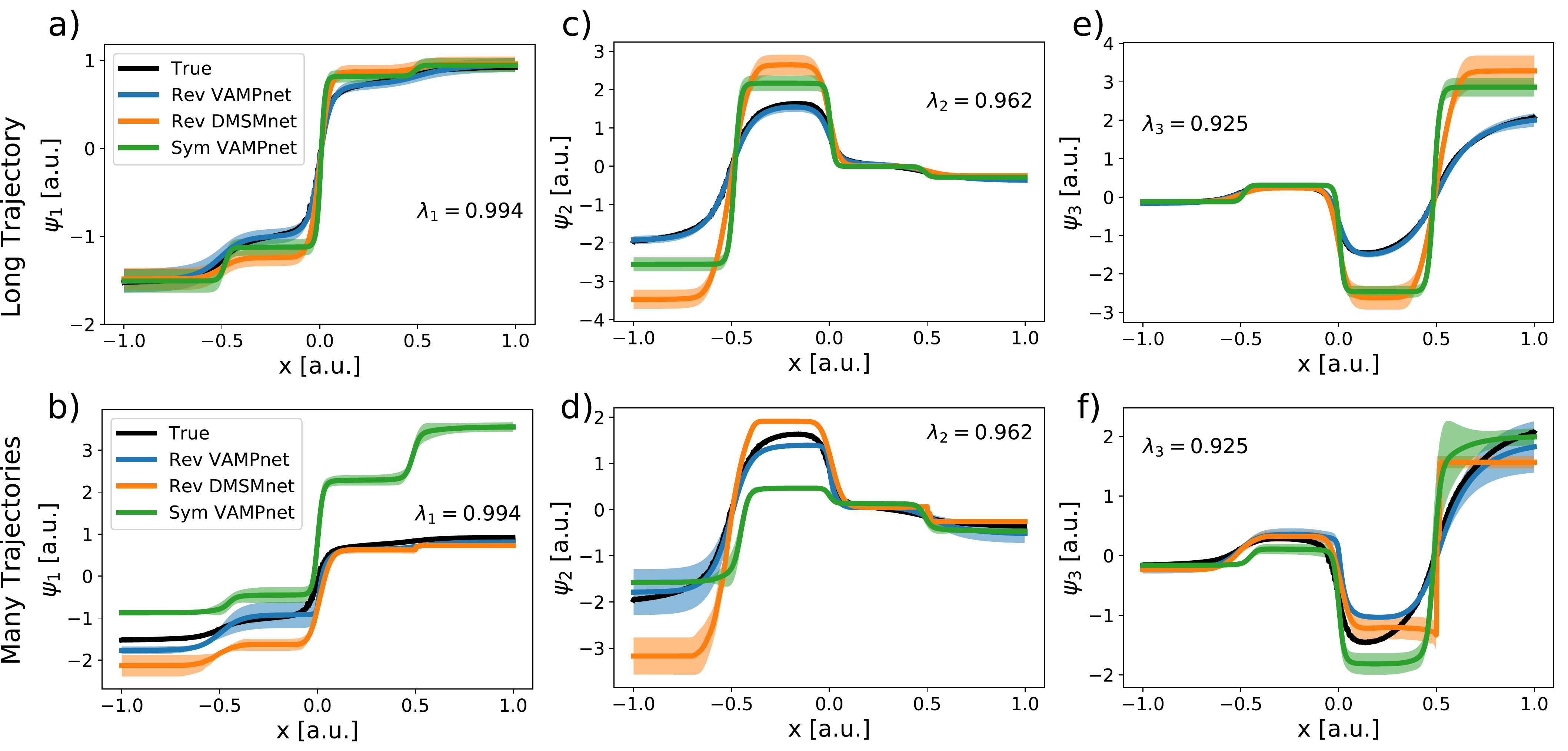}\centering

\caption{\label{fig:eigfunc_Prinz}\textbf{Estimating the three slowest eigenfunction
on the Prinz dataset with a RevVAMPnet, a SymVAMPnet, and a RevDMSM}.
a, c, e) Comparison of the eigenfunction estimated on one long trajectory.
b, d, f) Comparison for many short trajectories having an off-equilibrium
starting point distribution. Errors are estimated over 5 runs as two
sigma intervals.}
\end{figure}

Finally, we compare the approximation quality of the three slowest
non-trivial eigenfunctions for the Prinz potential for the three methods
(Fig. \ref{fig:eigfunc_Prinz}). The data from the analysis before
is reused of a single trajectory of length 50,000 frames, and of 10,000
trajectories 11 frames each. We compare the eigenfunctions against
a numerical approximation of the true eigenfunctions by a transition
matrix computed for a direct uniform 1,000-state discretization of
the $x$-axis for $2\cdot10^{7}$ frames as before \cite{PrinzEtAl_JCP10_MSM1}.

The RevVAMPnet is approximating accurately the true eigenfunctions
for all settings . In particular, it is able to recover the eigenfunctions
remarkably even in the case of the biased data. The SymVAMPnet results
are consistent with previous observations: the approximation of the
first two eigenfunctions of the non-equilibrium data (Fig. \ref{fig:eigfunc_Prinz}b,d)
are strongly biased, resulting in the underestimation of the implied
timescales. The constraints in the case of the RevDMSM lead to less
smoothly changing eigenfunctions and therefore less accurate approximations.
In the case of the long trajectory, both the SymVAMPnet and the RevDMSM
exhibit a stepwise behavior of the eigenfunctions, as they tend to
result in a harder assignment of states $\boldsymbol{\chi}$. Note
that for SymVAMPnet this can be alleviated by avoiding a SoftMax clustering
in the last layer and rather directly mapping onto the eigenfunctions
\cite{Chen_2019}.

\section{Conclusion}

We have introduced an end-to-end deep learning framework for molecular
kinetics that allows us to learn high-quality Markov models with physical
constraints such as reversibility and non-negativity of the learned
transition matrix. The proposed method is generally applicable for
reversible/non-reversible Markov State and Koopman models depending
on which constraints are enforced, thus it can be seen as an extension
and generalization of previous models such as VAMPnets and DeepMSMs.
Additionally, the optimization for the state classification and the
reversible transition matrix are not explicitly separate processing
steps compared to \cite{Chen_2019}. The proposed method is able to
estimate dynamical and stationary properties even from highly biased
data and gives state of the art results when studying the slow processes
and stationary characteristics of a small toy model. We also proved
that our model is a universal approximator of reversible Markov Processes
and fulfills the reversibility property with respect to the estimated
normalized stationary distribution.

Despite these advantages, a remaining concern is the optimization
procedure, which requires a good balance when fitting the three trainable
units at the same time. However, we are confident that the used protocol
of first fixing $\boldsymbol{\chi}$ and resetting $\mathbf{u}$ to
optimal values according to a non-reversible Koopman model during
the training process establishes a reproduceable procedure.

Furthermore, we expect that the maximum likelihood formulation of
the proposed method allows us to develop deep learning variants of
multi-ensemble MSMs (\cite{WuEtAL_PNAS16_TRAM,WuMeyRostaNoe_JCP14_dTRAM,ChoderaEtAl_JCP11_DynamicalReweighting,PrinzEtAl_JCP11_Reweighting,RostaHummer_DHAM,MeyWuNoe_xTRAM})
that alleviate rare event sampling, and augmented MSMs \cite{OlssonEtAl_PNAS17_AugmentedMarkovModels}
that incorporate experimental data into the model estimation.

\subsubsection*{}

\section{Acknowledgements}

This work was funded by the European Research Commission (ERC CoG
\textquotedblleft ScaleCell\textquotedblright ), Deutsche Forschungsgemeinschaft
(CRC 1114/A04, Transregio 186/A12, NO 825/4-- 1, Dynlon P8), and
the \textquotedblleft 1000-Talent Program of Young Scientists in China\textquotedblright .
Part of this research was performed while the author was visiting
the Institute for Pure and Applied Mathematics (IPAM), which is supported
by the National Science Foundation (Grant No. DMS-1440415).

\bibliography{all,hwu,own,references}

\section{Appendix}

\subsection{Proof of normalization\label{subsec:Proof-of-normalization}}

We show that the dynamical model (\ref{eq:Markovdensity}) has a normalized
transition and equilibrium density. By defining
\[
\mu(\mathbf{x})=\mathbf{u}^{T}\boldsymbol{\chi}(\mathbf{x})\rho_{1}(\mathbf{x}),
\]
we can obtain that
\begin{align*}
\int p_{\tau}(\mathbf{x},\mathbf{y})\mathrm{d}\mathbf{y} & =\int\boldsymbol{\chi}(\mathbf{x})^{T}\mathbf{S}\boldsymbol{\chi}(\mathbf{y})\boldsymbol{\chi}(\mathbf{y})^{T}\mathbf{u}\rho_{1}(\mathbf{y})\mathrm{d}\mathbf{y}\\
 & =\boldsymbol{\chi}(\mathbf{x})^{T}\mathbf{S}\mathbf{C}_{\tau\tau}^{\prime}\mathbf{u}=1\\
\int\mu(\mathbf{x})\mathrm{d}\mathbf{x} & =\mathbf{u}^{T}\bar{\boldsymbol{\chi}}=1
\end{align*}

\subsection{Proof of proposition 1\label{subsec:Proof-of-proposition1}}

\textit{Proof.} Since $\{\mathbf{x}_{t}\}$ is reversible, its transition
density can be decomposed as
\[
p_{\tau}(\mathbf{x},\mathbf{y})=\sum_{i=1}^{\infty}\lambda_{i}\varphi_{i}(\mathbf{x})\varphi_{i}(\mathbf{y})\mu(\mathbf{y}),
\]
where $\varphi_{i}$ is the eigenfunction corresponding to the eigenvalue
$\lambda_{i},$ $\{\varphi_{1},\varphi_{2},...\}$ is an orthonormal
basis of the Hilbert space $\{f|\langle f,f\rangle_{\mu}<\infty\}$
defined by the inner product
\[
\langle f,g\rangle_{\mu}=\int f(\mathbf{x})g(\mathbf{x})\mu(\mathbf{x})\mathrm{d\mathbf{x}},
\]
and $(\lambda_{1},\varphi_{1})=(1,\mu)$.

Considering a (d+1)-dimensional model
\begin{align*}
\boldsymbol{\chi}(\mathbf{x})^{T} & =(\boldsymbol{\varphi}(\mathbf{x})^{T},w_{1}(\mathbf{x})),\\
\mathbb{\mathbf{S}} & =\left[\begin{array}{cc}
\mathbb{\boldsymbol{\Lambda}}\\
 & 0
\end{array}\right],\\
\mathbf{\mathbf{u}} & =(0,...,0,1)^{T},
\end{align*}
we can obtain its transition density as
\begin{align*}
\hat{p}_{\tau}(\mathbf{x},\mathbf{y}) & =\boldsymbol{\chi}(\mathbf{x})^{T}\mathbf{S}\boldsymbol{\chi}(\mathbf{y})\rho_{1}(\mathbf{y})\boldsymbol{\chi}(\mathbf{y})^{T}\mathbf{u}\\
 & =\boldsymbol{\varphi}(\mathbf{x})^{T}\boldsymbol{\Lambda}\boldsymbol{\varphi}(\mathbf{y})\mu(\mathbf{y}),
\end{align*}
where $\boldsymbol{\varphi}=(\varphi_{1},...,\varphi_{d})^{T},$ $\boldsymbol{\Lambda}=\mathrm{diag}(\lambda_{1},...,\lambda_{d})$
and $w_{1}(\mathbf{x})=\rho_{1}(\mathbf{x})^{-1}\mu(\mathbf{x})$.

Then the Koopman operator $\mathcal{\hat{K_{\tau}}}$ deduced from
this model satisfies
\begin{align*}
||\mathcal{\hat{K_{\tau}^{*}}-\mathcal{K_{\tau}}||_{\mathrm{HS}}^{\mathrm{2}}} & =\int\int\rho_{0}(\mathbf{x})\frac{(\hat{p}_{\tau}(\mathbf{x},\mathbf{y})-p_{\tau}(\mathbf{x},\mathbf{y}))^{2}}{\rho_{1}(\mathbf{y})}\mathrm{d\mathbf{x}d\mathbf{y}}\\
 & =\int\int\rho_{0}(\mathbf{x})\left(\sum_{i=d+1}^{\infty}\lambda_{i}\varphi_{i}(\mathbf{x})\varphi_{i}(\mathbf{y})\right)^{2}w_{1}(\mathbf{y})\mu(\mathbf{y})\mathrm{d\mathbf{x}d\mathbf{y}}\\
 & \le C_{1}\int\rho_{0}(\mathbf{x})\sum_{i=d+1}^{\infty}\lambda_{i}^{2}\varphi_{i}(\mathbf{x})^{2}\mathrm{d\mathbf{x}}\\
 & \le C_{0}C_{1}\int\mu(\mathbf{x})\sum_{i=d+1}^{\infty}\lambda_{i}^{2}\varphi_{i}(\mathbf{x})^{2}\mathrm{d\mathbf{x}}\\
 & =C_{0}C_{1}\sum_{i=d+1}^{\infty}\lambda_{i}^{2},
\end{align*}
and the approximation error of the optimal (d+1)-dimensional model
is also bounded by $C_{0}C_{1}\sum_{i=d+1}^{\infty}\lambda_{i}^{2}.$

\subsection{Proof of proposition 2\label{subsec:Proof-of-proposition2}}

\textit{Proof.} Since $\{\mathbf{x}_{t}\}$ is reversible, $\mathcal{K_{\tau}}$
can be decomposed as
\[
\mathcal{K_{\tau}}f=\sum_{i=1}^{\infty}\lambda_{i}\langle f,\varphi_{i}\rangle_{\mu}\varphi_{i}
\]
with $\lambda_{1}=1$ and $\varphi_{1}\equiv1$. According to the
VAMP theory and considering that the proposed model is a specific
case of 
\[
\hat{p}_{\tau}(\mathbf{x},\mathbf{y})=\boldsymbol{\chi}(\mathbf{x})^{T}\mathbf{S}\boldsymbol{\chi}(\mathbf{y})\mu(\mathbf{y}),
\]

Eq. \ref{eq:vampe} can be proven.

If $\boldsymbol{\varphi}=(\varphi_{1},...,\varphi_{d})^{T}$ can be
represented by linear combinations of $\boldsymbol{\chi}(\mathbf{x})$,
i.e., there is an invertible matrix $\mathbf{R}=[R_{ij}]\in\mathbb{R}^{d\times d}$
so that
\[
\boldsymbol{\varphi}=\mathbf{R}\boldsymbol{\chi}.
\]
We can then construct the following model
\begin{align*}
\hat{p}_{\tau}(\mathbf{x},\mathbf{y}) & =\boldsymbol{\chi}(\mathbf{x})^{T}\mathbf{S}\boldsymbol{\chi}(\mathbf{y})\boldsymbol{\chi}(\mathbf{y})^{T}\mathbf{u}\mu(\mathbf{y})\\
 & =\boldsymbol{\varphi}(\mathbf{x})^{T}\mathbf{R}^{-T}\mathbf{SR}^{-1}\boldsymbol{\varphi}(\mathbf{y})\boldsymbol{\varphi}(\mathbf{y})^{T}\mathbf{R}^{-T}\mathbf{u}\mu(\mathbf{y})\\
 & =\boldsymbol{\varphi}(\mathbf{x})^{T}\boldsymbol{\Lambda}\boldsymbol{\varphi}(\mathbf{y})\mu(\mathbf{y}),
\end{align*}
where
\begin{align*}
\mathbf{S} & =\mathbf{R}^{T}\boldsymbol{\Lambda}\mathbf{R},\\
\mathbf{u} & =\mathbf{R}^{T}(1,0,....,0)^{T},
\end{align*}
and $\boldsymbol{\Lambda}=\mathrm{diag}(\lambda_{1},...,\lambda_{d})$.
It can be verified that the VAMP-E score of this model is equal to
$\sum_{i=1}^{d}\lambda_{i}^{2}$.

\subsection{Relationships to traditional Markov State Models\label{subsec:Relationships-to-traditional}}

Since in traditional MSMs the state definition $\boldsymbol{\chi}$
usually consists of a set of indicator functions partitioning the
full state space into Markov States, we show how in that case, for
the right choices of $\mathbf{u}$ and $\mathbf{S}$, our model obtains
the expected stationary and transition distribution.

Let $\boldsymbol{\chi}$ be the indicator functions of Markov states,
$\boldsymbol{\pi}$ be the stationary distribution vector, $\boldsymbol{\pi}_{\rho}=\bar{\boldsymbol{\chi}}$
the empirical distribution vector of states, and
\begin{align*}
\mathbf{u} & =\boldsymbol{\Pi}_{\rho}^{-1}\boldsymbol{\pi},\\
\mathbf{S} & =\mathbf{P}\boldsymbol{\Pi}^{-1},
\end{align*}
where $\boldsymbol{\Pi}=\mathrm{diag}(\boldsymbol{\pi}),$$\boldsymbol{\Pi}_{\rho}=\mathrm{diag}(\boldsymbol{\pi}_{\rho})$,
and $\mathbf{P}$ is the transition matrix. We can then express $\mu(\mathbf{x})$
as:
\begin{align*}
\mu(\mathbf{x}) & =\mathbf{u}^{T}\boldsymbol{\chi}(\mathbf{x})\rho(\mathbf{x})\\
 & =\boldsymbol{\pi}^{T}\boldsymbol{\Pi}_{\rho}^{-1}\boldsymbol{\chi}(\mathbf{x})\rho(\mathbf{x}),
\end{align*}
and
\begin{align*}
p_{\tau}(\mathbf{x},\mathbf{y}) & =\boldsymbol{\chi}(\mathbf{x})^{T}\mathbf{P}\boldsymbol{\Pi}^{-1}\boldsymbol{\chi}(\mathbf{y})\rho(\mathbf{y})\boldsymbol{\chi}(\mathbf{y})^{T}\boldsymbol{\Pi}_{\rho}^{-1}\boldsymbol{\pi}\\
 & =\boldsymbol{\chi}(\mathbf{x})^{T}\mathbf{P}\boldsymbol{\Pi}_{\rho}^{-1}\boldsymbol{\chi}(\mathbf{y})\rho(\mathbf{y})\boldsymbol{\chi}(\mathbf{y})^{T}\boldsymbol{\Pi}^{-1}\boldsymbol{\pi}\\
 & =\boldsymbol{\chi}(\mathbf{x})^{T}\mathbf{P}\boldsymbol{\Pi}_{\rho}^{-1}\boldsymbol{\chi}(\mathbf{y})\rho(\mathbf{y}).
\end{align*}

\end{document}